\documentclass[aps,prl,
  twocolumn,floatfix,
  superscriptaddress]{revtex4-1}

\usepackage{graphicx}
\usepackage{color}

\begin{document}

\title{Solitary versus Shock Wave Acceleration in Laser-Plasma Interactions}
\author{Andrea Macchi}\email{andrea.macchi@ino.it}
\affiliation{Istituto Nazionale di Ottica, Consiglio Nazionale delle Ricerche, Research Unit ``Adriano Gozzini'', Pisa, Italy}
\affiliation{Dipartimento di Fisica ``Enrico Fermi'', Universit\`a di Pisa, Largo B. Pontecorvo 3, I-56127 Pisa, Italy}
\author{Amritpal Singh Nindrayog}
\author{Francesco Pegoraro}
\affiliation{Dipartimento di Fisica ``Enrico Fermi'', Universit\`a di Pisa, Largo B. Pontecorvo 3, I-56127 Pisa, Italy}
\affiliation{Istituto Nazionale di Ottica, Consiglio Nazionale delle Ricerche, Research Unit ``Adriano Gozzini'', Pisa, Italy}

\date{\today}

\begin{abstract}
The excitation of nonlinear electrostatic waves, such as shock and solitons, 
by ultraintense laser interaction with overdense plasmas and related ion 
acceleration are investigated by numerical simulations. 
Stability of solitons and formation of shock waves is strongly dependent
on the velocity distribution of ions.
Monoenergetic components in ion spectra are produced by ``pulsed'' reflection 
from solitary waves. Possible relevance to recent experiments on ``shock 
acceleration'' is discussed.
\end{abstract}

\maketitle

In the interaction of superintense ($I=10^{18}-10^{21}~\mbox{W cm}^{-2}$) 
laser pulses with overdense plasmas (i.e. having electron density 
$n_e>n_c$, the cut-off density), the light pressure ranges from Gigabar to
Terabar values and, like a piston, may sweep out and compress the 
laser-produced plasma pushing its surface at nearly relativistic speeds. 
Such combination of strong compression and acceleration is often described
as the generation of strong shock waves (briefly, ``shocks'')
propagating towards the bulk of the plasma. 
These light-pressure driven shocks may cause heating of solid targets, 
as experimentally inferred \cite{akliPRL08} and be of interest for the 
production of warm dense matter. 
In addition, shocks may lead to acceleration of
ions if the latter are reflected by the shock front as a moving wall. 
In moderately overdense 
and hot plasmas, where the shock waves are of 
collisionless nature, shock acceleration may lead to higher ion energies 
than the widely studied target normal sheath acceleration (TNSA) mechanism 
\cite{wilksPoP01}, 
as suggested on the basis of numerical simulations \cite{silvaPRL04}. 
Recently, two experiments employing hydrogen gas jet targets and CO$_2$ 
lasers reported on monoenergetic proton beams produced via shock 
acceleration \cite{palmerPRL11,haberbergerNP11}.  
The study of laser-driven shocks
in the laboratory may also help the understanding of similar processes in
astrophysical environments \cite[and references therein]{martinsApJ09}

The above mentioned simple picture of ion acceleration indicates that, 
as far as the 
shock velocity $v_s$ is constant, the reflected ions should have velocity
$2v_s$ and produce a monoenergetic peak in the spectrum. \textcite{silvaPRL04} 
report that such peak would evolve into a spectral plateau
due to further acceleration in the sheath field at the rear side of the target, 
so that suppression of such field (for instance by producing a smooth density 
gradient at the rear side) could preserve monoenergeticity. However, in the
present Letter we show that obtaining monoenergetic spectra is not 
straightforward, even neglecting the side effect of the sheath field. 
We observe monoenergetic
peaks in the simulations only when short-duration ion bunches are accelerated
by \emph{solitary} waves generated by the laser-plasma interaction. 
Moreover, the formation and the evolution of both soliton- or shock-like waves
is highly dependent on the velocity distribution of background ions. 
A crucial point is that, for a ion distribution without a velocity spread, 
either \emph{all} the ions or \emph{none} of them may be reflected from a 
moving electrostatic field front, depending on the ratio 
$\Phi_{\mbox{\tiny max}}/v_s^2$ where $\Phi_{\mbox{\tiny max}}$ is the potential jump 
at the front. In the first case, a strong loading of the moving structure 
occurs, causing deceleration and collapse of the shock or soliton and a 
consequent broadening of the reflected ions spectrum.

Before describing our simulation results we briefly recall some key aspects
of the fluid theory of collisionless electrostatic shocks and solitons 
\cite{tidman-krall}.
The most relevant point
for what follows is that the existence of reflected ions is \emph{integral}
to collisionless shocks and not a secondary effect; it may be used as a
signature of ``true'' shock formation in the simulations. 
When reflected ions are absent, solutions corresponding to electrostatic
\emph{solitons} are found. Hence, a necessary and general condition for such 
solitons to exist with a velocity $v_s$ is that the electrostatic potential 
energy jump $e\Phi$ has a peak value 
\begin{equation}
e\Phi_{\mbox{\tiny max}}<m_iv_s^2/2 \, ,
\label{eq:threshold}
\end{equation}
so that background ions are not reflected by the soliton. 
Within the fluid theory with the electrons in an isothermal Boltzmann 
equilibrium \cite{tidman-krall} at the temperature $T_e$, the condition on 
the potential poses an upper limit on the Mach number 
 $M=v_s/c_s<1.6$, where $c_s=\sqrt{T_e/m_i}$ is the speed of sound.
The other condition is that the soliton must be supersonic, i.e. $M>1$.

\begin{figure*}
\includegraphics[width=0.98\textwidth]{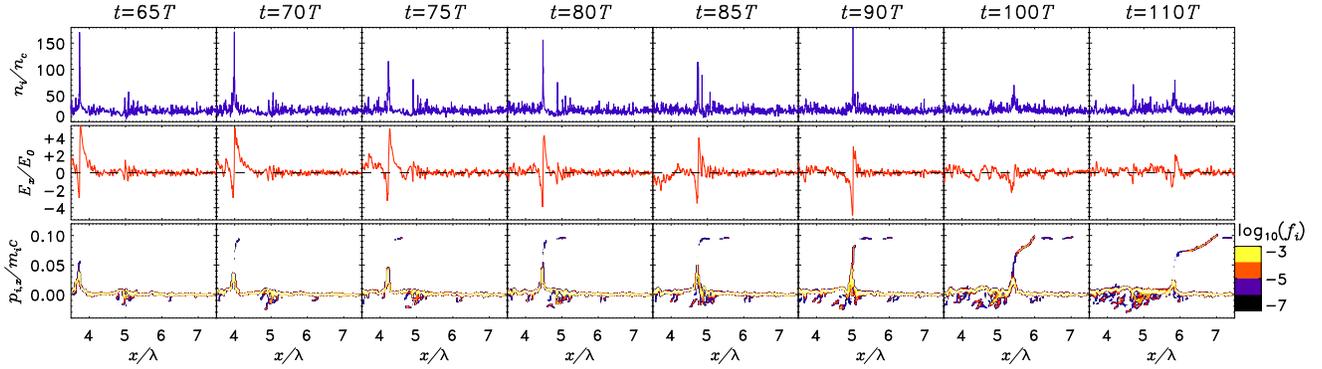}
\caption{(Color online). Snapshots of the evolution of a Solitary Acoustic Wave 
(SAW) at eight different times. The upper, middle and bottom rows show
the ion density $n_i$, the electrostatic field $E_x$ 
and the contours of the
$f_i(x,p_x)$ ion phase space distribution in a $\log_{10}$-scale. 
The laser pulse impinges from the left reaching the plasma boundary 
($x=0\lambda$)
at $t=0T$. 
Simulation parameters are $a_0=16$, $\tau=4T$ and $n_e=20n_c$. 
$n_i$ is normalized to $n_c$, $E_x$ to $E_0=m_e\omega c/e$ and $p_x$ to $m_pc$.}
\label{fig:SAWacceleration}
\end{figure*}

Since we focus on basic aspects we restrict to one-dimensional Particle-In-Cell 
simulations for the sake of simplicity and high numerical resolution.
We found that the numerical results converged only for sufficiently
high values of the number of particles per cell $N_p$, although qualitatively 
similar features were observed also for lower $N_p$. This suggests that 
kinetic effects play an essential role and thus low-density tails in the
distribution functions must be resolved accurately. In the simulations
reported below,  $N_p=800$ and the 
spatial and temporal resolution $\Delta x=c\Delta t=\lambda/400$, where
$\lambda$ is the laser wavelength. 

We now analyze a representative simulation with the following set-up.
The laser pulse was linearly polarized with a peak amplitude $a_0=16$, 
where $a_0=0.85(I\lambda^2/10^{18}~\mbox{W cm}^{-2}\mu\mbox{m}^2)^{1/2}$,
and duration $\tau=4T$ (FWHM), with $T$ the laser period; the temporal profile 
is composed by $1T$ long, $sin^2$-like rising and falling ramps 
and a $3T$ plateau. 
The plasma had a slab, square-like profile with initial ion and electron 
densities $n_i=n_e=n_0=20n_c$ and $15\lambda$ thickness. Ions were protons
($Z=A=1$).
For reference, the laser pulse front reaches the front surface placed
at $x=0$ at the time $t=0$. 

Eight snapshots of the profiles of ion density $n_i$ and longitudinal 
electrostatic field $E_x$ and of the $f_i(x,p_x)$ phase space distribution 
of ions are shown in Fig.\ref{fig:SAWacceleration}. 
In the early stage of the interaction, the laser pulse accelerates a fraction 
of strongly relativistic electrons with energy of several 
$m_ec^2$ which penetrate into the target
and later recirculate across it, driving heating of bulk electrons. 
A solitary structure is generated at the front surface under the action of the 
laser pulse and then propagates into the plasma bulk at a constant velocity
$v_s \simeq 0.05c$. 
This value is close to the ``hole boring'' velocity $v_{\mbox{\tiny hb}}$
\cite{macchiPRL05,robinsonPPCF09a} which in our case is given 
by
\begin{equation}
v_{\mbox{\tiny hb}}
=a_0c\left(\frac{1+R}{2}\frac{Z}{A}\frac{m_e}{m_p}\frac{n_c}{n_e}\right)^{1/2}
\label{eq:vHB}
\end{equation}
where the reflection coefficient is measured to be $R \simeq 0.75$ in the
simulation of Fig.\ref{fig:SAWacceleration}, yielding 
$v_{\mbox{\tiny hb}} \simeq 0.06c$. 
At $t=65T$, the solitary structure is located at $x\simeq 3.7\lambda$
(first frame of Fig.\ref{fig:SAWacceleration}). 
The ion density has a very strong spike, reaching values
up to $~\simeq 9$ times the background density. 
The electric field around the density spike has a sawtooth shape. 
For reference we call the structure we observe a SAW (for Solitary Acoustic 
Wave). 
The snapshots of SAW density and field profiles 
are qualitatively resemblant to those of electrostatic
solitons as described in Ref.\cite{tidman-krall}.
However, several of the dynamics features we describe below are not described
by standard fluid theory.

\begin{figure}[b]
\includegraphics[width=0.48\textwidth]{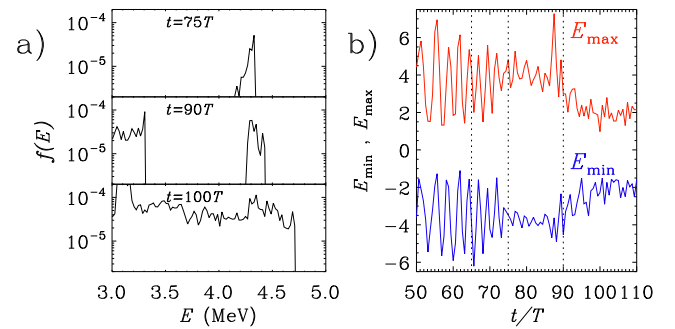}
\caption{(Color online). 
a):
ion spectra from the simulation of Fig.\ref{fig:SAWacceleration} 
at the times $t=75T$, $90T$ and $100T$. Only ions in the same spatial range as 
Fig.\ref{fig:SAWacceleration}, i.e. $3.5<x<7.5$, are included in the spectra. 
b): temporal evolution of the maximum (red line) and the minimum 
value (blue line) of the electric field for the SAW structure observed in 
Fig.\ref{fig:SAWacceleration}. The vertical dashed lines mark the instants 
$t=65T$, $75T$ and $90T$
at which ion reflection events start as seen in Fig.\ref{fig:SAWacceleration}.
}
\label{fig:spectrum3}\label{fig:SAWoscillation}
\end{figure}

Numerical integration of $E_x$ over $x$ for the SAW at $t=65T$ 
(first snapshot in Fig.\ref{fig:SAWacceleration}) yields 
$e\Phi_{\mbox{\tiny max}} \simeq 0.78E_0\lambda \simeq 4.9m_ec^2$. 
Since $m_iv_s^2/2 \simeq 2.3m_ec^2<e\Phi_{\mbox{\tiny max}}$
one would expect ion reflection to occur promptly from the SAW field.
This occurs indeed in the simulation, producing a bunch of high-energy
ions with very small momentum spread as can be observed at 
$t=70T$ and $t=75T$. However, violation of the condition (\ref{eq:threshold})
and consequent ion reflection do not destroy the SAW; at $t=75T$, the SAW 
is still present with almost unperturbed velocity, and ion reflection has 
stopped. Between $t=75T$ and $t=85T$ a second bunch is formed in a very
similar way, as observed between the third and fifth snapshot in 
 Fig.\ref{fig:SAWacceleration}. Both bunches correspond to a monoenergetic
high-energy peak in the spectrum, as shown in Fig.\ref{fig:spectrum3}~a);
the FWHM of the peak at $t=75T$ is less than $1\%$.

The acceleration of a third bunch at $t=90T$ is correlated with a 
much stronger perturbation of the SAW whose amplitude and velocity decrease.
As a consequence, the high-energy part of the spectrum broadens up, and 
monoenergetic features are lost as seen in Fig.\ref{fig:spectrum3}~a) 
at $t=100T$.

The apparent behavior of the SAW is related to the observation that the 
electric field amplitude is not constant in time, as for the ``standard''
soliton solution, but it \emph{oscillates} as shown in 
Fig.\ref{fig:SAWoscillation}. The temporal behavior of the maximum and minimum
values shows that the electron cloud around the ion density spike
oscillates back and forth. 
The oscillation is quenched after the generation of the second fast bunch at 
$t \simeq 75T$. At this instant, the potential jump at the SAW front is 
$e\Phi_{\mbox{\tiny max}} \simeq 0.5E_0\lambda \simeq 3.1m_ec^2$, slightly
above the stability threshold (\ref{eq:threshold}).
Eventually, the overall amplitude strongly decreases after the generation of
the third bunch, as also observed in the last frames in 
Fig.\ref{fig:SAWacceleration}.
Apparently, such ``collapse'' of the SAW occurs after a ``collision'' with a 
slower, counterpropagating structure that can also be noticed in the frames of 
Fig.\ref{fig:SAWacceleration}.

The above reported observations led us to infer that the local velocity
distribution of ions has a crucial role in determining both the stability of 
solitary waves and the possibility to generate ``true'' shocks whose signature
would be, in the framework of fluid theory, the formation of a continuous flow
of reflected ions. If the ions are ``cold'', i.e. have no energy spread, 
reflection from a moving potential barrier may occur either for none or for 
all of the ions. In the latter case, the wave may quickly 
lose its energy by accelerating the whole bulk of ions. 

\begin{figure}[t]
\includegraphics[width=0.48\textwidth]{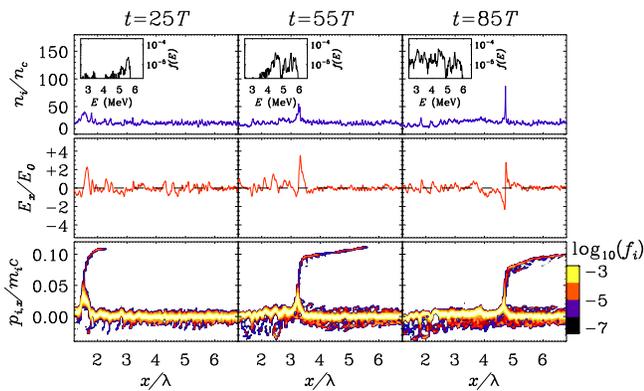}
\caption{(Color online). Snapshots at three different times of
the $n_i$, $E_x$ and $f_i(x,p_x)$ for a simulation identical to that of 
Fig.\ref{fig:SAWacceleration} but for an initial ion temperature 
$T_i \simeq 10^{-2}m_ec^2=5~\mbox{keV}$, showing the onset of ``continuous'' ion
reflection. The scales are the same as in Fig.\ref{fig:SAWacceleration} to show
that the perturbation in $n_i$ and $E_x$ has a lower value in the present case.
The insets show the corresponding ion spectra, limited to the ions 
in the region around the shock and excluding ions accelerated at the target
boundaries.
}
\label{fig:warm}\label{fig:spectrum3warm}
\end{figure}

In order for a collisionless shock to form, it must be possible for the 
wave to ``pick up'' from the ion distribution only a fraction of the ions 
with the right energy to be reflected. This can occur in a plasma with 
energy spread, i.e. warm ions. 
Fig.\ref{fig:warm} shows results of a simulation with identical 
parameters as Fig.\ref{fig:SAWacceleration}, but the initial ion temperature
$T_i=5~\mbox{keV}$. Formation of a shock-like structure with steady
reflection of ions is observed. The shock front velocity is found to decrease 
in time 
which may be interpreted as due to
the wave energy transfer into reflected ions. As a consequence, 
with respect to the ``cold'' ions case narrow monoenergetic peaks almost never
appear at any time in the spectrum (see the insets in 
Fig.\ref{fig:spectrum3warm}) and
the high-energy tail progressively broadens towards lower energies, 
although the cut-off energy is a bit higher with respect to the ``cold'' 
case. Thus, in this case, a spectral ``plateau'' is produced because of
shock deceleration, rather than due to further acceleration of the reflected 
ions in the rear side sheath. 
Some modulations are apparent both in the phase space density and
in the spectral plateau of reflected ions; they may
be related to the oscillation of the electric field at the wave front, similar
to Fig.\ref{fig:SAWoscillation}.

The effect of the background ion distribution can be also evidenced 
studying a case with parameters identical to Fig.\ref{fig:SAWacceleration} but
a shorter length of the plasma ($3\lambda$). In this case, not shown here
for briefity, the SAW generated at the front side of the plasma 
eventually reaches the rear side and propagates in the expanding sheath
until it reaches the region in which the ion 
velocity is such that the ions are now reflected by the SAW potential 
(since the ions in the sheath move in the same direction of the SAW, their 
velocity in the rest frame of the SAW is lower than in the ``laboratory'' 
frame, hence they are reflected more easily).
From this point on the SAW quickly loses its energy in accelerating ions and 
collapses. A similar mechanism was also described by \textcite{zhidkovPRL02b}.

When the laser pulse duration is increased with respect to the above reported
simulations, we observe multiple peaks of the ion density 
and sawtooth spatial oscillations of the electric field. 
Although a structure showing multiple oscillations behind a front is 
reminiscent of a collisionless shock wave as seen in textbooks, again we do 
not observe in general a steady ion reflection at the front.
The density peaks move at different velocities and thus disperse in time. 
Hence the structure may be interpreted as a multi-peak SAW,
generated due to pulsed hole boring acceleration 
\cite{macchiPRL05} at a rate which is approximately the same for simulations 
having same pulse intensity and plasma density, so that a longer pulse 
duration allows acceleration of a sequence of ion bunches. 

\begin{figure}
\includegraphics[width=0.48\textwidth]{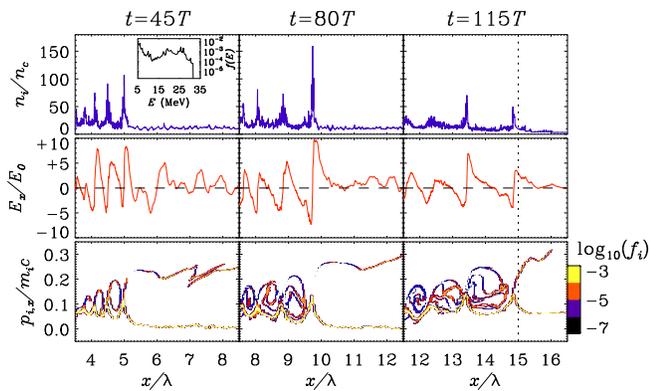}
\caption{(Color online). 
Snapshots of multi-peak structures at three different times from a 
``long pulse'' simulation. 
The inset shows the spectrum at $t=45T$, limited to the ions in the same
region as the plots.
The vertical dashed line marks the initial position of the rear side of the
target.
Simulation parameters are $a_0=16$, $\tau=65T$ and $n_e=10n_c$.
Units are the same as in Fig.\ref{fig:SAWacceleration}.
}
\label{fig:multiSAW}\label{fig:silva}
\end{figure}

A representative ``long pulse'' simulation is shown in Fig.\ref{fig:silva}.
The laser pulse has peak amplitude $a_0=16$ (like the above reported 
``short pulse'' simulations) with $5T$ rise and fall ramps and a $60T$ plateau.
The plasma density and thickness are $10n_c$ and $15\lambda$, respectively.
These parameters are close to those of previously reported 1D simulations 
(see Fig.1 in Ref.\cite{silvaPRL04}) 
and indeed some features observed in the ion phase space  
and density profiles look very similar. However, in our simulations the higher 
number of computational particles allows us to highlight additional details
in the phase space distribution of 
Fig.\ref{fig:silva} (bottom row) such as vortex structures behind the
front, corresponding to ``trapped'' ions bouncing between adjacent peaks, where
a potential well is formed. 
The observation of ions trapped in the multi-peak, nonlinear structure of the
electric field and the quite broad distribution along the momentum axis of the
most energetic ions suggest that those latter may be actually accelerated by 
``surfing'' the longitudinal wave structure, rather than being merely reflected 
by the field at the front. 
In addition, we observe significant oscillations of the electric field 
\emph{ahead} of the wave front (i.e. of the rightmost density peak in 
Fig.\ref{fig:silva}), which may have been excited by fast electrons and 
are related to modulations in the momenta of the highest energy ions. The
spectra of the latter is quite broad (inset of Fig.\ref{fig:silva} even 
neglecting the contribution of TNSA at the rear side.

Our study suggests that generating highly monoenergetic ions by
``shock acceleration'' is not straightforward, as in our simulation 
we observe narrow spectra only as resulting from SAW ``pulsations'' and
as a transient effect, since as shown in 
Figs.\ref{fig:SAWacceleration} and \ref{fig:SAWoscillation} the SAW collapse
ultimately produces a broad spectrum masking the monoenergetic peak. 
Producing a monoenergetic spectrum might not coexist with efficiency,
because the reflection of a large fraction of ions by the moving structure
(either a shock or a soliton) would ultimately cause a strong loading effect
decreasing its field and velocity. This is in qualitative agreement with
the experimental results of \textcite{haberbergerNP11} where the number of ions 
($\sim 2.5 \times 10^5$) in the narrow spectral peak at $\sim 22~\mbox{MeV}$ 
implies a conversion efficiency $<10^{-8}$ 
of the $60~\mbox{J}$ pulse energy. A direct comparison of our simulations with
the experiment in Ref.\cite{haberbergerNP11}, which involves several additional
issues such as e.g. the pulse shape and the plasma profile, is beyond the scope 
of the present Letter.

All of the above described effects related to SAW generation, dynamics and
acceleration of ions in the \emph{bulk}
are observed only with linear polarization, 
for which electrons are heated up to relativistic energies. 
For circular polarization, no ion acceleration occurs in the bulk. 
Although \textcite{palmerPRL11} report on 
``protons accelerated by a radiation pressure driven shock'', we find that 
the experiment can be better described as evidence of hole boring (HB)
acceleration \cite{macchiPRL05,robinsonPPCF09a}. Nevertheless, the two 
different mechanisms might produce similar ion spectra 
because the velocity of the shock launched at the surface is of the 
order of $v_{\mbox{\tiny hb}}$ [Eq.(\ref{eq:vHB})],
while the hole boring
process produces via wave breaking ion bunches of velocity 
$v_i\simeq 2v_{hb}$ \cite{macchiPRL05}
as if the ions were ``reflected'' from the plasma surface as a piston driven 
by light pressure. 

In conclusion, in our simulations we observe the laser-driven generation
of nonlinear, collisionless acoustic waves having either a solitary or a 
multipeak structure depending on the laser pulse duration. 
In particular we observe a ``pulsation'' of solitary waves leading to the
generation of small numbers of monoenergetic ions. 
In a cold ion background, wave loading effects prevent ``true'' shock wave 
formation and efficient monoenergetic acceleration. In general the dynamics of
``shock acceleration'' in the plasma bulk appears to be more complex than the
simple picture of ``reflection from a moving wall''.

\acknowledgments
This work has been supported by the Italian Ministry for University and Research
(MIUR) through the FIRB ``Futuro in Ricerca'' project 
``Superintense Laser-Driven Ion Sources''.

\hyphenation{Post-Script Sprin-ger}

\end{document}